\documentclass[12pt,amsfonts]{article}
\usepackage{amsfonts}
\def\one{1\hskip-.37em 1}

\def\ts{\textstyle}

\def\hpi{{\hat\pi}}
\def\mfh{\mathfrak h}
\def\half{\textstyle{\frac{1}{2}}}

\def\mfH{\mathfrak{H}}
\def\threebytwo{\textstyle{\frac{3}{2}}}

\def\p{\phi}

\def\k{\kappa}

\def\l{\lambda}

\def\t{\textstyle}

\def\ra{\rightarrow}
\def\tint{{\textstyle\int}}

\def\hp{{\hat\pi}}
\def\hph{{\hat\phi}}
\def\s{\hskip.08em}

\def\d{\partial}

\def\b{\begin{eqnarray*}}  
\def\e{\end{eqnarray*}}    
\def\bn{\begin{eqnarray}}  
\def\en{\end{eqnarray}}   

\def\<{\langle}
\def\>{\rangle}

\def\no{\nonumber}

\def\hk{\hat{\kappa}}
\def\{{\lbrace}
\def\}{\rbrace}
\title{Quantum Field Theory \\With No Zero-Point Energy}  
\author{John R. Klauder\footnote{john.klauder@gmail.com}\\
Department of Physics and Department of Mathematics\\
University of Florida,
Gainesville, FL 32611-8440}
\begin{document}
\maketitle
\begin{abstract} Traditional quantum field theory can lead to enormous zero-point energy, which markedly  disagrees with experiment. Unfortunately, this situation is
built into conventional canonical quantization procedures. For identical classical theories, an alternative quantization procedure, called affine field quantization, leads to the desirable feature of
having a vanishing zero-point energy. This procedure
has been applied to renormalizable and nonrenormalizable covariant scalar fields, fermion fields, as well as general relativity. Simpler models are offered as an introduction to affine field quantization. \end{abstract}

\section{Introduction}
The source of zero-point energy for a free scalar field 
is readily canceled by normal ordering of the Hamiltonian operator, which leaves every term with an annihilation operator. Unfortunately, for a non-free scalar field,
say with a quartic potential term, normal ordering cannot eliminate the zero-point energy since there is always a term composed only of creation operators.

This situation is inevitable with canonical quantization (CQ) when a classical field $\p(x)\ra \hph(x)=[A(x)^\dag+A(x)]/\sqrt{2}$,
 where $x\in\mathbb{R}^s, s\in\{1,2,3,\ldots\}$, and its momentum field $\pi(x)\ra\hp(x)=i\hbar[A(x)^\dag-A(x)]/\sqrt{2}$, $[A(x),A(y)^\dag]=\delta(x-y)\one$, and $[\hph(x),\hp(y)]=i\hbar\s\delta(x-y)\one$.  As usual, the ground state of the Hamiltonian,
designated by $|0\>$, ideally is the
 unique, normalized, `no particle' state for which $A(x)\s|0\>\equiv 0$ for all $x$, and Hilbert space is given by suitable combinations of terms such as $A(x_1)^\dag\s A(x_2)^\dag\s\cdots\s A(x_N)^\dag\s|0\>$ for all $N<\infty$ when smeared with suitable elements $f(x_1, \ldots, x_N)\in L^2(\mathbb{R}^{sN})$.  

 However, there is also a different approach. In place of the momentum, we substitute the {\it dilation field} $\k(x)\equiv\pi(x)\p(x)$---a
  field that is featured in the enhanced quantization (EQ) program \cite{E3}---which leads to the
  Poisson bracket $\{\p(x),\k(y)\}=\delta(x-y)\s\p(x)$, a formal Lie algebra for the affine group. As the analog of a current commutation relation \cite{ZZ}, an affine quantization of these variables leads to $\p(x)\ra \hph(x)\equiv
  \tint B(x,\l)^\dag\l\s B(x,\l)\,d\l$ and $\k(x)\ra \hk(x)\equiv \tint B(x,\l)^\dag \tau(\l) B(x,\l)\,d\l$, where $\tau(\l)\equiv-\half i\hbar[\l(\d/\d\l)+
  (\d/\d\l)\l]$. Here $B(x,\l)\equiv A(x,\l)+c(\l)\one$, where $\l\in\mathbb{R}$, $A(x,\l)|0\>=0$ for all $(x,\l)$, and $c(\l)$ is the real `model function'; for the example offered in Sec.~3
  $c(\l)=|\l|^{-1/2}\s W(\l)$, where $0<W(\l)=W(-\l)<\infty$ with $0<\epsilon<W(0)$. Local operator products are given 
  by an operator product expansion \cite{ZZ}, which leads to renormalized (R) products such as
 $\hph(x)^2_R=b\tint B(x,\l)^\dag\s\l^2\s B(x,\l)\,d\l$, etc.; here, the fixed factor $b\propto(\rm {length})^{-s}$ serves to maintain  proper dimensions.
 Such local products become terms in the quantum Hamiltonian, which is given by
  \bn &&\mfH\equiv \tint B(x,\l)^\dag\s\mfh(\d/\d\l,\l)\s B(x,\l)\,d\l\,d^s\!x \no\\
  &&\hskip1.09em =\tint A(x,\l)^\dag\s\mfh(\d/\d\l,\l)\s A(x,\l)\,d\l\,d^s\!x\;, \en
which requires  that $\mfh(\d/\d\l,\l)\s c(\l)=0$. For the example of Sec.~3, which has a classical Hamiltonian density of $\half \pi(x)^2+ V(\p(x))$, for some $V$ with no spatial derivatives,  then, with  $\hbar=1$, \bn && \mfh(\d/\d\l,\l)=-c(\l)^{-1}\d/\d\l \s c(\l)^2\s  \d/\d\l c(\l)^{-1} \no \\ &&\hskip5.01em = -\d^2/\d\l^2+c(\l)^{-1}\d^2 c(\l)/\d\l^2\label{88}\;. \en We require that $\tint [\l^2/(1+\l^2)]c(\l)^2\,d\l<\infty$, and, to obtain a unique ground state, we insist that $
 \tint c(\l)^2\,d\l=\infty$, which ensures that $\mfh>0$, as normalized expectations of the top line of (\ref{88}) would confirm.

 Hence: (i) $\mfH\ge0$, (ii) $\mfH$ has a {\it unique ground state} $|0\>$ (because $c(\l)\not\in L^2(\mathbb{R})$ implies that all eigenvalues of $\mfh$ with normalizable eigenfunctions are positive),
 and (iii) $\mfH$ has {\it no zero-point energy} (because {\it all} terms in $\mfH$ lead with an annihilation operator). The Hilbert space has combinations of $A(x_1,\l_1)^\dag A(x_2,\l_2)^\dag\cdots A(x_N,\l_N)^\dag\s|0\>$ for all $N<\infty$ when smeared with suitable elements $f(x_1, \l_1, \ldots, x_N,\l_N)\in L^2(\mathbb{R}^{(s+1)N})$.

\section{Nonrenormalizable Models Exposed} 
  A simple `toy' example can help understand nonrenormalizable models. Consider the one-variable problem with a classical action functional given by
  $I_g=\tint_0^T [\s\half\dot{q}(t)^2 -\half q(t)^2-g q(t)^{-4}\s]\,dt$, where the coupling constant $g\ge0$. The free model, with $g=0$, has a domain of functions $D_{g=0}=\{q:\,\tint_0^T[\dot{q}^2+q^2]\s dt<\infty\}$, while the interacting theory, with any $g>0$, has a domain of functions
  $D_{g>0}=\{q: \,\tint_0^T[\dot{q}^2+q^2 +q^{-4}]\s dt<\infty\}$. Clearly, $ D_{g>0}\subset D_{g=0}$, and, in particular, while solutions to the
  equations of motion within $D_{g=0}$
  cross back and forth over $q=0$, solutions to the equations of motion within $D_{g>0}$ {\it never cross $q=0$ and are always positive or always negative}. In fact, solutions to the equations of motion when $g>0$
  {\it find that the limiting behavior of such solutions as $g\ra0$ is continuously connected to a solution that belongs to $D_{g>0}$ and does {\it not} belong to $D_{g=0}$}. Specifically, a free model solution is given by $q(t)=B\s\cos(t+\beta)$, for any $B$ and $\beta$, while a $g\ra0$ solution is given by $q(t)=\pm\s |B\s\cos(t+\beta)|\ne0$. In brief, as $g\ra0$, {\it an interacting model is continuously connected with the free action functional but with the constraint that its associated domain is $D_{g>0}$}. Thus, the `interacting models' are {\bf not} continuously connected to the `free model'! We say that the $g\ra0$ solutions belong to a {\it pseudofree model} rather than the {\it free model}.

  Quantum mechanically, the propagator for the free model is given by
    \bn K_f(q'',T;q',0)=\Sigma_{n=0,1,2,3,\ldots}\; h_n(q'')\s h_n(q')\,e^{-i(n+1/2)T/\hbar}\;, \en
    where $\{\s h_n(q)\s\}_{n=0}^\infty$ are Hermite functions. On the other hand, the propagator for the pseudofree model \cite{BS}
    is given by
    \bn &&K_{pf}(q'',T;q',0)=\lim_{g\ra0}\,{\cal N}_g\int_{q(0)=q'}^{q(T)=q''} e^{(i/\hbar)\tint_0^T[(\dot{q}^2-q^2)/2-g\s q^{-4}]\,dt}\,{\cal D}q\no\\
    &&\hskip7em =2\s\theta(q'' q')\s\s\Sigma_{n=1,3,5,7\ldots}\;h_n(q'')\s h_n(q')\,e^{-i(n+1/2)T/\hbar}\;, \en
    where $\theta(u)\equiv1$ if $u>0$ and $\theta(u)\equiv0$ if $u<0$. Clearly, the interaction term has changed the paths that contribute to the
    path integral. {\it This result implies that any attempt at a perturbation series must be taken about the pseudofree theory
    because if it is taken about the free theory every term will diverge!} {\it {\bf Domains matter!}}

  \section{Ultralocal Scalar Model: \\A Nonrenormalizable Example}
The classical Hamiltonian for this model is given by
 \bn H(\pi,\p)=\tint \{\s\half [\pi(x)^2  +m_0^2\s \p(x)^2\s]+g_0\s \p(x)^4\s \}\, d^s\!x \;, \label{kp} \en
 with the Poisson bracket $\{\p(x), \pi(x')\}=\delta(x-x')$. This model has also been discussed before, e.g., \cite{BCQ,ul,UL2}, and we will offer just an
overview of its quantization.

Let us first introduce an $s$-dimensional  spatial lattice for (\ref{kp}) of step $a>0$ to approximate
the  spatial integral, which leads to the expression for a regularized classical Hamiltonian given by
  \bn H_K(\pi,\p)={\ts\sum}_k [ \half (\s\pi_k^2+m_0^2\s\p_k^2\s)+g_0\s \p_k^4 ]\,  a^s\;, \en
  where $k=\{k_1,k_2,\ldots,k_s\}\in K$, $k_j\in\mathbb{Z} \equiv\{0,\pm1,\pm2,\ldots\}$, where $K$ is large but finite  and $a^s$ represents a cell volume. Clearly, a suitable limit of $K\ra \infty$ and $a\ra 0$ is understood such that $K\s a^s\ra V$, where $V\le\infty$ is the spatial volume implicit in (\ref{kp}).

  Next we discuss a quantization based on CQ for the lattice Hamiltonian prior to taking a spatial limit with $a\ra0$. Since there is no connection of the dynamics between spatial points, it follows that at each lattice site $\pi_k\ra a^{-s}\s P_k$ and $\p_k\ra Q_k$ and $H_k(\pi_k,\p_k)\ra {\cal{H}}_k(P_k,Q_k)$. Moreover,
  at each site ${\cal{H}}_k(P_k,Q_k)=\{\,\half[\s a^{-2s}\s P_k^2+m_0^2\s Q_k^2]+g_0\s Q_k^4 +{\cal{O}}(\hbar;P_k,Q_k; a)\,\}\, a^s$, where we assume the last term is polynomial in $P$ and $Q$ and vanishes if $\hbar\ra0$.
  Here the
   ground state $|\psi_{0;k}\>$ fulfills the relation ${\cal{H}}_k(P_k,Q_k)\, |\psi_{0;k}\>=0$.
 The normalized Schr\"odinger representation $\<\p_k|\psi_{0;k}\>$ 
  has the form $\exp[- Y(\p_k;\hbar,a)\s a^s/2]$, where $-\infty<Y(\p_k;\hbar,a)=Y(-\p_k;\hbar,a)<\infty$, and the characteristic function of the regularized overall ground-state distribution is given by
      \bn &&C_K(f)=\Pi_k\,\tint e^{i f_k \p_k\s a^s/\hbar}\,  e^{- Y(\p_k;\hbar,a)\s a^s}\,d\p_k/\tint  e^{- Y(\p_k;\hbar,a)\s a^s}\; d\p_k \no\\
     &&\hskip3.13em =\Pi_k\,\tint e^{i f_k \p_k\s }\,  e^{- Y(\p_k\hbar/ a^s;\hbar,a)\s a^s}\,d\p_k/\tint  e^{- Y(\p_k\hbar/ a^s;\hbar,a)\s a^s}\; d\p_k \;. \en
    The final issue involves taking the continuum limit as $a\ra0$. To ensure a reasonable limit it is helpful to expand the term involving $f_k$ in a
  power series, which leads to
     \bn C_K(f)=\Pi_k\{ 1-\t{\frac{1}{2!}} f_k^2\<
     \p_k^2\>_a+\t{\frac{1}{4!}}f_k^4\<\p_k^4\>_a-\t{\frac{1}{6!}}f_k^6\<\p_k^6\>_a-\ldots \}\;. \en
     In order to achieve a meaningful continuum limit, it is necessary that $\<\p_k^2\>_a\propto a^s$, which, in the present case,
     means that $\<\p_k^{2\s j}\>_a\propto a^{s\s j}$ (since the distribution is `tall and narrow' about $\p_k=0$), leading to the result that the continuum limit becomes
     $C(f)=\exp[-R\tint f(x)^2\,d^s\!x ]$ for some $R$, $0<R<\infty$.
     This result is a standard result of the Central Limit Theorem \cite{CLT}, which implies a Gaussian (= free) ground state for CQ.

     Next 
      we introduce classical affine fields $\p(x)$ and $\k(x)$, and
     the classical Hamiltonian becomes
       \bn H'(\k,\p)=\tint \{\s\half\s [\s\k(x)\s\p(x)^{-2}\s\k(x) + m_0^2\s\p(x)^2] +g_0\s \p(x)^4 \} \, d^s\!x \;. \en
       A quantum study begins by promoting the classical affine fields to operators $\k(x)\ra\hk(x)$ and $\p(x)\ra\hph(x)$, and
       it formally follows that
       \bn  \hk(x)\s\hph(x)^{-2}\s\hk(x)=\hpi(x)^2+{\t{\frac{3}{4}}}\s\hbar^2\s\delta(0)^2\s \hph(x)^{-2} \;.  \en
       On the same spatial lattice as before, the quantum Hamiltonian is chosen as
       \bn 
       {\cal{H}}'_K(\hk,\hph) 
        ={\ts\sum}_k \{\half [ \s-\s\s a^{-2s}\s\hbar^2 \d^2/\d\p_k^2\s + a^{-2s}\s F\s\hbar^2 \p_k^{-2} +m_0^2\s
        \p_k^2] +g_0\s \p_k^4\s \}\,a^s \,, \en
       where $F\equiv
       (\half-b\s a^s)(\threebytwo-b \s a^s)$ is a regularized version of $\t{\frac{3}{4}}$, and $b$ is a fixed constant so
        that $b\s a^{s}$ is dimensionless. The ground state $\psi_0(\p)$ satisfies ${\cal{H}}'_K(\hk,\hph)\s \psi_0(\p)=0$ and, for $ba^s\ll1$, has a form such that
       \bn \psi_0(\p)^2=\Pi_k\,(ba^s)\,e^{-Z(\p_k;\hbar,a)\s a^s}\,|\p_k|^{-(1-2ba^s)}\;. \label{pp} \en
        The exponent $Z$ prevents divergence of an integral of the ground-state distribution when $|\p_k|\gg1$, but when $|\p_k|\ll1$ an integral
        of the ground-state distribution is proportional to $(ba^s)^{-1}+{\cal O}(1)$, in which case the pre-factor $(ba^s)$,
        when very tiny, provides the overall normalization. 
            Once again we study the characteristic function
       \bn &&\hskip-2.5em C(f)=\lim_{a\ra0} \Pi_k\tint\s\{\s e^{if_k\p_k\s a^s/\hbar}\,(ba^s)\,e^{-Z(\p_k;\hbar,a)\s a^s}\,|\p_k|^{-(1-2ba^s)}\;d\p_k \s\}\no\\
           &&=\lim_{a\ra0}\Pi_k\{1-(ba^s)\tint[1-e^{if_k\p_k\s a^s/\hbar}] \,e^{-Z(\p_k;\hbar,a)\s a^s}\,|\p_k|^{-(1-2ba^s)}\;d\p_k \s\}\no\\
           &&=\exp\{-b\tint d^s\!x\,\tint[1-e^{if(x)\s\l/\hbar}]\, e^{-z(\l;\hbar)}\,d\l/|\l| \}\;, \label{uu} \en
           where $\l=\p_k\s a^s$, 
           and suitable factors in $Z$ undergo an operator product expansion renormalization which leads to $z$.
           Besides a Gaussian distribution found by CQ, the resultant form (\ref{uu}) 
            is  the only other result of the Central Limit Theorem, namely,  a (generalized) Poisson distribution (based on the new moments
            $\<\p_k^{2\s j}\>_a\propto a^s$ for all $j\in\{1, 2, 3,\ldots\}$). Moreover,
           the classical limit as $\hbar\ra0$ for this solution has been shown \cite{E3} to yield the starting classical Hamiltonian.

           As an example, suppose we take the limit $g_0\ra0$. In this case, $c(\l)=|\l|^{-1/2}\s e^{-b m \l^2/2\s\hbar}$, and thus
              \bn \mfh(\d/\d\l,\l)= \half\s b^{-1}\s[ \s-\hbar^2\s\d^2/\d\l^2+ \t{\frac{3}{4}}\s\hbar^2\s \l^{-2}+ b^2 m^2 \l^2\s] \;. \en
           The resultant characteristic function is given by
              \bn C_0(f)=\exp\{-b\tint d^s\!x\s\tint[1-e^{if(x)\s\l/\hbar}\s]\s e^{-b\s m\l^2/\hbar}\,d\l/|\l|\s \} \;, \label{rrr}\en
              which does not represent a traditional free model. However, the ground state implicitly described here reflects the fact that the
              domain of the classical free action functional is strictly larger than the domain of the classical interacting action functional and
             which is  unchanged in the limit of the coupling constant  $g_0\ra0$.
               Just as with the `toy' example in Sec.~2, and from a path integral viewpoint, this domain modification leads to a different
                set of paths than in the free-model path integral. Moreover, the spectrum of the relevant quantum Hamiltonian
                associated with (\ref{rrr}) has a uniform
                spacing $(2\hbar\s m)\s j$, where $j\in\{0,1,2,3,\cdots\}$, and which, as promised, correctly reflects a vanishing zero-point
                 energy \cite{BCQ}.

                 This last reference also discusses different singular terms that are ${\cal O}(\hbar^2)$, which may be relevant
                 in alternative applications.

 \section{Additional Studies using Affine Variables}
Enhanced quantization and affine variables have also been applied to other problems.
The use of affine variables to discuss idealized
 cosmological models \cite{Y1,WW} has led to gravitational bounces rather than an initial singularity of the universe. Further use of
 affine variables applied to idealized gravitational models has been given in seveal papers, e.g., \cite{T1,T2,T3}, and
 references therein. More complex applications
 include covariant scalar fields \cite{E3,E1,E2,PP,JK1},
 and quantum gravity \cite{E3,G1,G2,G3}, for which confirmation requires multiple selected computer simulations.

 \section*{Acknowledgements} The referee is thanked for several useful corrections and helpful suggestions.


\begin{thebibliography}{99}  

\bibitem{E3} J. R. Klauder, {\it Enhaced Quantization: Particles, Fields \& Gravity}, (World Scientfic, Singapore, 2015).

\bibitem{ZZ} C. Itzykson and J.-B. Zuber, {\it Quantum Field Theory}, (McGraw-Hill, New York, 1980).

\bibitem{BS} B. Simon, ``Quadratic Forms and Klauder's Phenomenon: A Remark on Very Singular Perturbations'', J. Funct. Anal. {\bf 14}, 295 (1973).
\bibitem{BCQ} J. R. Klauder, {\it Beyond Conventional Quantization} (Cambridge UniveJrsity Press, Cambridge, 2000).
\bibitem{ul}J. R. Klauder, ``Ultralocal Scalar Field Models", Commun. Math. Phys. {\bf 18}, 307-318 (1970).
\bibitem{UL2}J. R. Klauder,
 ``Ultralocal Quantum Field Theory", Acta Physica Austriaca, Suppl. {\bf VIII}, 227-276 (1971).
 \bibitem{CLT} https://en.wikipedia.org/wiki/Central\_limit\_theorem.


\bibitem{Y1} J. R. Klauder and E. W. Aslaksen, ``Elementary Model for Quantum Gravity", Phys. Rev. D {2}, 272 (1970).
\bibitem{WW} M. Fanuel and S. Zonetti, ``Affine Quantization and the Initial Cosmological Singularity'', EPL 101, 10001 (2013); arXiv:1203.4936v3.

\bibitem{T1}  H. Bergeron, E. Czuchry, J.-P. Gazeau, P. Ma\/lkiewicz, W. Piechocki, ``Singularity Avoidance in a Quantum
Model of the Mixmaster Universe'', Phys. Rev. D {92}, 124018 (2015); arXiv:1501.07871: ``Smooth Quantum Dynamics of Mixmaster
Universe'', Phys. Rev. D { 92}, 061302 (2015); arXiv:1501.02174.
\bibitem{T2}  H. Bergeron, E. Czuchry, J.-P. Gazeau, P. Ma\/lkiewicz, ``Nonadiabatic Bounce and an Inflationary Phase in
the Quantum Mixmaster Universe'', Phys. Rev. D { 93}, 124053 (2016); arXiv:1511.05790.
\bibitem{T3}  H. Bergeron, E. Czuchry, J.-P. Gazeau, P. Ma\/lkiewicz,  ``Vibronic  Framework  for  Quantum  Mixmaster
Universe'', Phys. Rev. D { 93}, 064080 (2016); arXiv:1512.00304.
\bibitem{E1}J. R. Klauder, ``Enhanced Quantization: A Primer'', J. Phys. A: Math. Theor. {\bf 45},  285304 (8pp) (2012); arXiv:1204.2870.
\bibitem{E2} J. R. Klauder, ``Enhanced Quantum Procedures that Resolve Difficult Problems'', Rev. Math. Phys. 27 (2015) no.05, 1530002 (43pp):
 arXiv:1206.4017.

\bibitem{PP} J. R. Klauder, ``Nontrivial Quantization of $\phi^4_n,\, n\ge2$'', Theor. Math. Phys. 182 (2015) no.1, 83-89;
Teor. Mat. Fiz. 182 (2014) no.1, 103-111; arXiv:1405.0332.
\bibitem{JK1} J. R. Klauder, ``Scalar Field Quantization Without Divergences In All Spacetime Dimensions'',
J. Phys. A: Math. Theor. 44, 273001 (30pp) (2011); arXiv:1101.1706.

 \bibitem{G1} J. R. Klauder, ``Noncanonical Quantization of Gravity. I. Foundations of Affine Quantum Gravity'', J. Math. Phys. 40, 5860 (1999);
 arXiv:gr-qc/9906013.
 \bibitem{G2} J. R. Klauder, ``Noncanonical Quantization of Gravity. II. Constraints and the Physical Hilbert Space'', J. Math. Phys. 42, 4440 (2001);
  arXiv:gr-qc/0102041.
  \bibitem{G3} J. R. Klauder, ``Recent Results Regarding Affine Quantum Gravity'', J. Math. Phys. 53, 082501 (2012); arXiv:1203.0691.






\end{thebibliography}
\end{document}